# Construction of AC Motor Controllers for NOvA Experiment Upgrades

Patrick Cooley

Fermi National Accelerator Laboratory

Community College Initiative

August 4, 2011

Saint Louis Community College at Florissant Valley

Linda Purcell-Taylor

Abstract

I have been constructing Alternating Current (AC) motor controllers for manipulation of particle beam detectors. The capability and reliability of these motor controllers are essential to the Laboratory's mission of accurate analysis of the particle beam's position. The device is moved in and out of the beam's path by the motor controller followed by the Neutrinos at the Main Injector Off-Axis $v_e$ Appearance (NOvA) Experiment further down the beam pipe. In total, I built and tested ten ac motor controllers for new beam operations in the NOvA experiment. These units will prove to be durable and provide extremely accurate beam placement for NOvA Experiment far into the future.

Introduction

The purpose of my work in this area is to provide more capability for the Neutrinos at the Main Injector Off-Axis $v_e$ Appearance (NOvA) Experiment. The NOvA experiment hopes to observe neutrino ($v_e$) oscillations, ascertain various neutrino masses and determine symmetry between neutrinos and antineutrinos. This experiment will attempt to further explain the evolution and distribution of galaxies throughout the cosmos. NOvA consists of a 200-ton detector on the Fermi National Accelerator site and a 15,000 ton detector in Soudan, Minnesota. Neutrinos that reach the Fermi detector must pass through the earth before emerging at the Soudan detector. NOvA uses a liquid scintillator to detect neutrino interactions. The charged particles produced by the neutrino interaction inside the detector cause the liquid scintillator to produce light that is captured by optical fibers and carried to light-sensitive detectors [1]. As particle beams travel down a beam pipe at Fermi National Accelerator Laboratory, measurements of the beam's exact position are necessary for proper placement and intensity magnitude. Minimization of beam loss is critical for experimental integrity. The AC Motor Controller moves Segmented Wire Ionization Chambers, Wheel Targets, Secondary Emission Monitors and Collimators [2] in and out of the beam line, which is accomplished remotely and offers read back and detector position control. Another feature of the motor controller is DC Braking, which allows for finer control of the various motor systems that are used. The four-channel AC motor controllers consist of three power supplies, a transformer, a bridge rectifier, four relay boards, a display board, and a nine-computer card tray (two per channel and one indicator board). This setup includes the cabinet and the attendant read back, motor connectors, control connectors, fuses, dc brake switches and timing sync coaxial connector. The Controllers were exclusively designed by Daniel Schoo of Fermi National Accelerator Laboratory [3], [4], [5], [6]. The AC Motor Controller can control four separate units independently and remotely using the Accelerator's Control Network (ACNET) and the central computer system in the Main Control Room of the Accelerator Division. ACNET is a system by which the particle beam profile can be monitored and the various units can be moved in and out of the beam. ACNET also provides the means by which to see a visual profile of the beam in real time (Figure 1), that aids in proper beam placement. Control of the units can also be achieved manually. There are two field replaceable cards for each channel. The first, called the "Limit and Readback" board, contains relays for the position sensing switches and read back electronics for position sensing. Every functional channel uses one Limit and Readback board regardless of the type of device controlled. The second card needed is a standard or customized board used to control a particular device depending on how it is expected to move. Control signals from the interface boards activate relays controlling the desired direction of movement. There are four relays for each channel. Two relays control the clockwise direction of the motor and two relays control the counterclockwise direction. The AC Motor Controller is designed such that either a three wire or four wire motor can be operated by connecting the motor leads to the appropriate connector pins. The indicator board performs three functions. First, it provides visual indicators that each of the four power supplies is operating.

Second, it receives timing pulses that optically isolate, buffer and distribute pulses throughout the card cage backplane to each of the interface boards. Third, the circuit also includes a pulse detector and lamp driver which flash a light emitting diode to indicate the presence of a pulse. The NOvA experiment requires ten new AC Motor Controllers for upgrading beam line operations. I have constructed, tested and repaired ten of these units, which required wire tracing, soldering and verifying proper power distribution throughout the controllers.

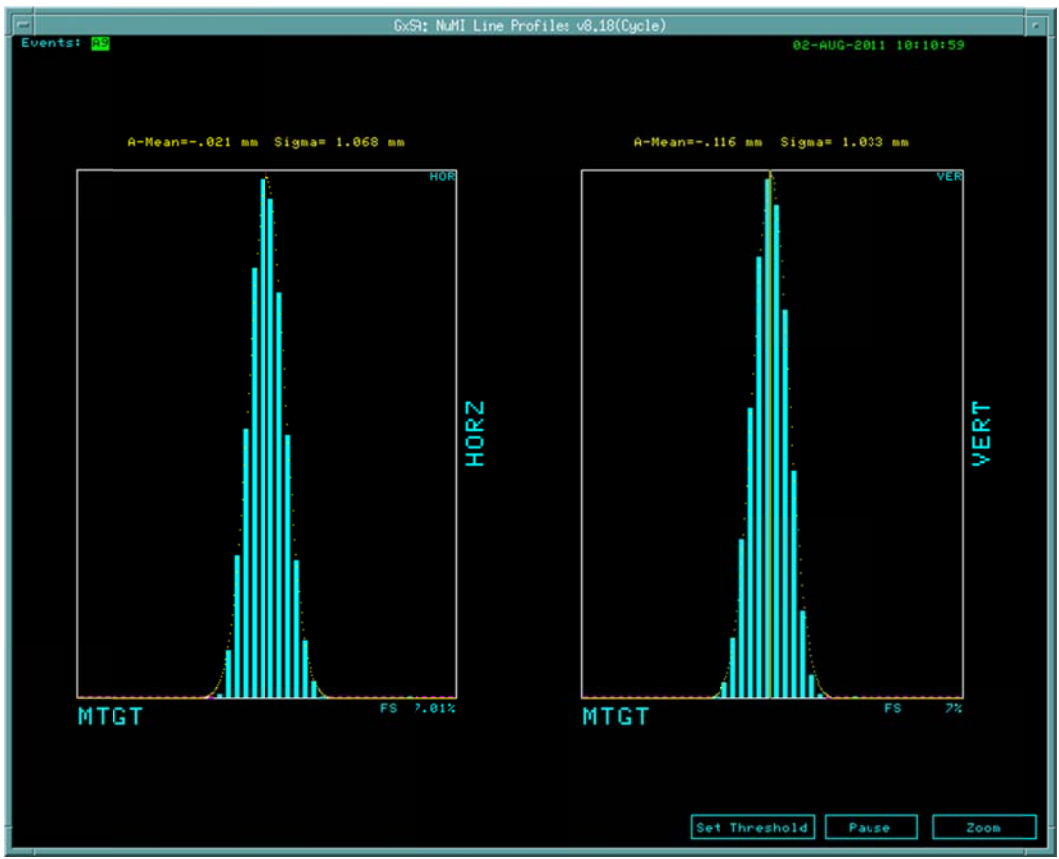

Figure 1 NoVa Target Multiwire Beam Profile

Methods and Materials

The construction of AC Motor Controllers required the extensive use of various schematics of components within the units. Data sheets of discrete components were also needed to give specific operating parameters and expected characteristics. Electrical signal tracing methods were applied throughout the entire process. Current flow and wiring diagrams were drawn up to offer quick analyses of the full circuit at any time during the construction and testing phases. An operational AC Motor Controller was also utilized for comparison and for testing individual components such as display boards. Testing equipment designed specifically for this equipment was employed as well as industry standard equipment [7]. Daniel Schoo has designed and manufactured devices to test the various motor controls, the read back and ribbon cables. Motor drive simulators were also designed by Mr. Schoo to assess the controllers on three and four wire motors and the limit switches of each channel. The ribbon cables' continuity was analyzed by a component that systematically pulsed through each wire to ensure connectivity of a thirty-four pin connector. A potentiometer connected to the read back fifteen pin connector and a multimeter were used, allowing for voltage verification. Motor control was ensured using this method as well. The industry standards, Fluke multimeter, the Weller Soldering Iron and the Tektronix digital storage Oscilloscope were employed in this construction project as well as many other standard tools. Drills, punches, and tap and die equipment were operated for specialized holes needed. Schematics, created by Dan Schoo (Figures 2, 3), allowed for analysis of specific portions of the overall layout. There were specific schematics for connections of the controller to different detectors to allow for checking various parameters. An actual detector, the Secondary Emissions Monitor, was also in use for testing the final product.

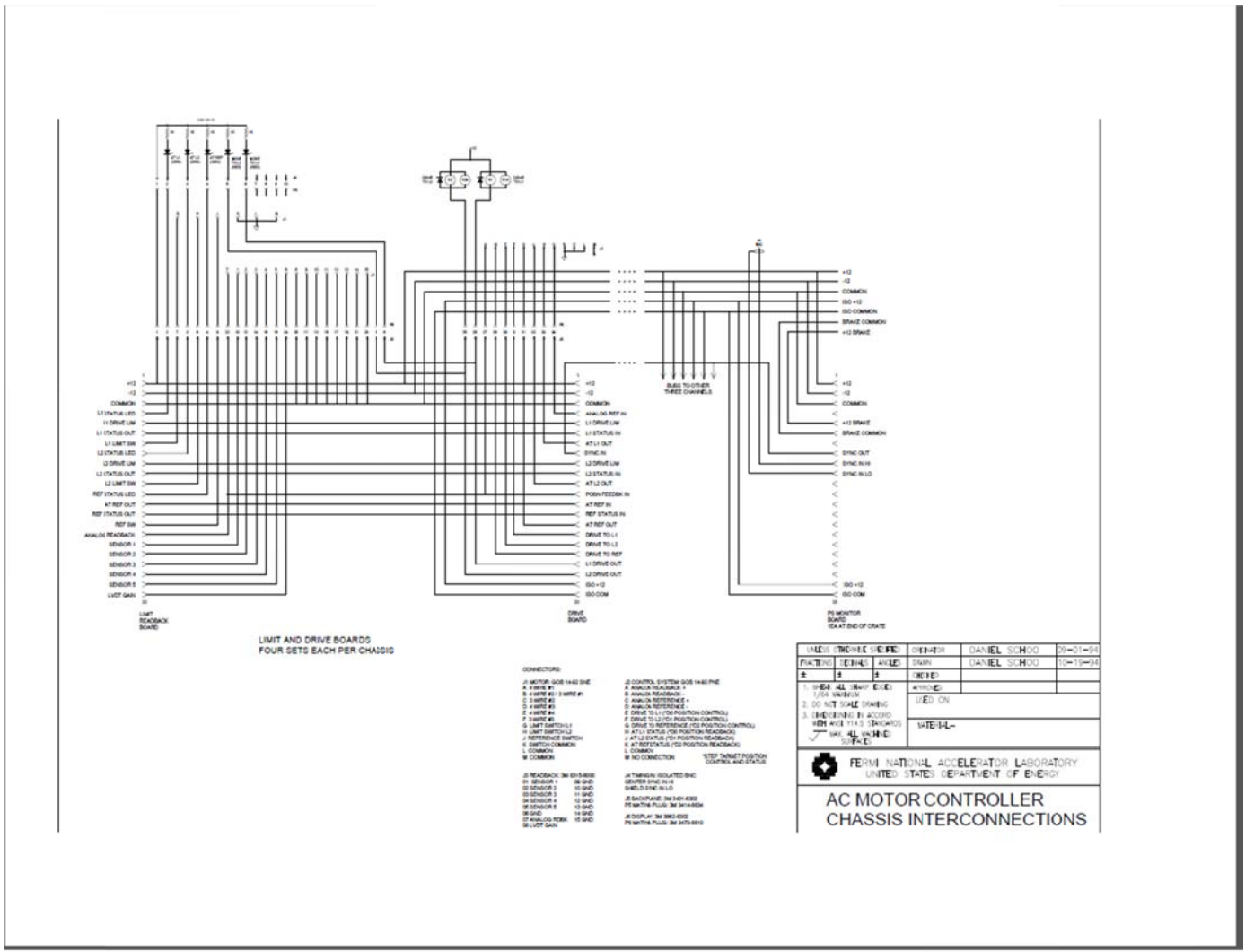

Figure 2 Chassis Schematic

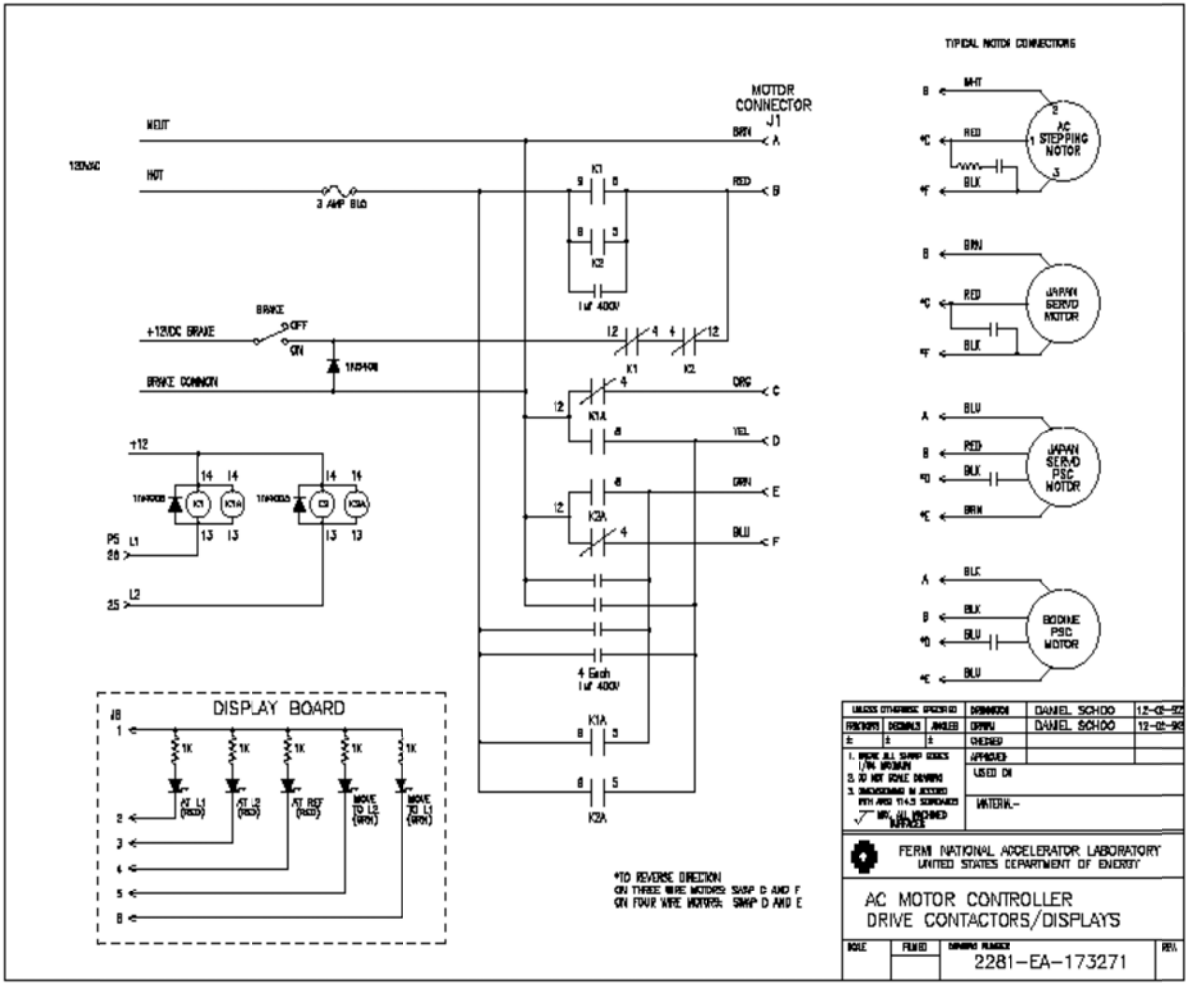

**Figure 3 Drive Board Schematic**

Results

Although there were relatively few difficulties in the construction of the AC Motor Controllers, certain complications were certainly vexing. I had problems with the controllers when they were first energized. I initially troubleshot the main board and the internal connections. I used the current flow diagram and determined there was no problem with the chassis wiring from the ac line through the transformer, the bridge rectifier and through the power supplies. Power was also evident in the main control board. I traced the signals via the schematic and located the problem. The trouble was apparently in the relay boards. I found a problem with the ac power lead to the relays. (Pin B to the K1- Normally Open from the Normally Closed of the first relay). I also had a problem with one relay board that was working but had too little power when energized; I ended up finding a reversed diode. I also had a problem with another diode which allowed signal in one direction but not the other. This diode was also reversed. Another problem solved was an improper bridge rectifier was installed. After confirming the proper rectifier and its operational parameters, the correct bridge rectifier was installed. On one controller, there was a wrong pin assignment. On yet another unit it seems one of the power supplies and the main board was not operating at all. Prompt replacement of said items enabled proper operation. All of the units encountered problems with the main display boards, improper placement of a resistor was determined to be the problem and quickly remedied. The card holder board of one unit was found to be defective during a test of read back and control connectors. This malfunction required the replacement and soldering of a new card holder backplane board. The sub-assembly of each component was accomplished by third parties and thus had to be thoroughly verified. Full operation was ensured by the testing of single capabilities individually and then combinations of various operations.

Discussion

The proper operational capacity of the ten AC Motor Controllers was fully tested and connected to a Secondary Emissions Monitor. In the future these control units could possibly be built cheaper with the use of smaller and more operationally capable components. Usage of discrete components is needed to isolate signals, to ensure that there is no interference to the operation of a single channel, when multiple channels are functioning. Advancements in switching and timing components might still offer signal isolation and proper operation with smaller or cheaper components. The integration of microprocessors or field programmable gate arrays could also offer some of this flexibility. Since the original manufacture, advances in integrated chip design and signal logic have progressed exponentially. These advancements may offer more attributes and functionality. Durability is certainly an asset in the current model and as such this feature should never be compromised. Another important characteristic of the AC Motor Controller is almost any functional component can be replaced while still in the field. Replacing an individual card is all a technician has to do to have the controller operate a different type of detector. These facets must be retained in any future design as well. It is necessary for the AC Motor Controller to be placed in an experimental environment for years without maintenance or upkeep. The success of the Laboratory's mission requires that undue breakdowns in essential equipment be minimized. Thus any new design considerations should always retain the original component's robust design.

## Acknowledgements

This project was conducted at Fermi National Accelerator Laboratory in the Instrumentation Section of the Accelerator Division. The construction of the ten AC Motor Controllers was accomplished during the period of May through August of 2011. My mentor, Linda Purcell-Taylor, Senior Electronics Technician, provided extensive assistance and guidance throughout the entire assembly process. The construction was also aided by Daniel Schoo, Electronic

Engineer and Rick Pierce, Senior Electronics Technician.  These three individuals were very supportive and instructive in my development and my understanding of the necessary processes.  I want to thank the United States Department of Energy, the Office of Science, the Community College Institute of Science and Technology and the Fermi National Accelerator Laboratory for this wonderful and enlightening program.  The entire program with lectures, tours and instruction will surely aid in my future endeavors and professional growth.  Funding for this initiative is crucial to the future of technology in our great nation and is greatly appreciated.